# The SPACE of AI: Real-World Lessons on AI's Impact on Developers

Brian Houck, Travis Lowdermilk, Cody Beyer, Steven Clarke, Ben Hanrahan

The rise of artificial intelligence (AI) in software development has sparked intense debate, especially among executives and engineering leaders eager to understand its impact. Some believe AI will soon replace entire categories of developer work, while others dismiss it as an overhyped distraction. But amid all the speculation, one perspective is often missing: that of the developers who are actually using these tools.

On one side of the coin, it seems clear that developers themselves are unworried about the potential of AI to replace them in the foreseeable future. Only 10% of software engineers express concern that AI might take their jobs[1]. For those working with these tools every day, the reality is clear: AI tools aren't replacing developers; they're augmenting them, helping them work faster, smarter, and with less toil. On the flip side, some organizations remain hesitant about whether AI is worth the investment.

The truth, assuredly, lies somewhere in the middle. While developers see clear benefits from AI, many organizations struggle to assess its true impact. Most efforts to measure AI's value focus narrowly on how much faster developers can complete coding tasks[2]. This misses the bigger picture: speed is just one piece of the puzzle. Productivity is multidimensional, as captured in the SPACE framework, which considers not just Activity and Efficiency but also factors like Satisfaction, Collaboration, and delivering meaningful Performance[3]. Not-to-mention that developers only spend 14% of their time coding new features, a statistic that underscores how incomplete a focus on coding speed alone really is.[4]

To better understand how AI is reshaping the developer experience, we conducted a study that asked developers how AI tools are affecting their work. Unlike other studies, which often focus on a single tool like GitHub Copilot, this research considers a broader range of AI tools. By combining quantitative insights with qualitative user stories, we aim to explore not just how developers are using AI but also how it impacts their productivity across the dimensions of SPACE.

This paper will first outline the study design, then dive into the findings, examining AI adoption trends, its influence on individual and team productivity, and the ways it is, or isn't, shaping the broader developer experience. Finally, we'll discuss actionable recommendations for teams and organizations looking to harness AI effectively. Through this work, we hope to shed light on AI's true value in the developer's toolkit and uncover what organizations can do to maximize its potential.

## Methodology

To explore how artificial intelligence (AI) is impacting developer productivity and satisfaction across the dimensions of SPACE, we conducted a mixed-methods research study combining survey data with qualitative insights derived from interviews and focus groups.

**Survey Design**

The survey consisted of a mix of multiple-choice questions, Likert-scale ratings, and open-text prompts. The core questions aimed to assess how AI tools impact developers' experiences across the SPACE dimensions, such as Satisfaction, Performance, Activity, Collaboration, and Efficiency. Additionally, respondents were asked about team-level adoption, the availability of AI training resources, and their own frequency of AI use. The survey was optimized for brevity, with an average completion time of seven minutes.

**Survey Participants**

The survey was distributed to 3,500 individuals, from which we received 530 responses, yielding a response rate of 15%, which is in line with similar studies. Over 80% of the study participants were from Microsoft, but there were developers from at least 15 other companies, including Airbnb, Atlassian, Charles Schwab, DoorDash, JetBrains, Meta, Netflix, Reddit, and Yelp. Non-Microsoft participants were recruited via professional networks and industry contacts, with company names provided on an optional basis.

The Microsoft sample was drawn from a randomized pool of developers within Microsoft, filtered to include individual contributors in the U.S., excluding those in Finance, Legal, or HR. Individuals who had received survey invitations within the past 12 months or explicitly opted out of research participation were also excluded. The external sample spanned industries, company sizes, and geographies, providing a broader context for comparison. We acknowledge that these constraints, particularly the geographic restriction, may influence the results.

**Survey Execution**

The survey was conducted in August of 2024, using Microsoft Forms. Participants were invited via email and incentivized with a chance to win a $50 gift card. To ensure privacy,

responses were collected anonymously, and no identifying information was linked to the results.

**Survey Limitations**

As with any survey-based research, this study has limitations. Response bias may influence the results, particularly among external participants who represent a smaller sample size. Additionally, while the survey captured perceptions of AI's impact, it does not directly measure productivity improvements or satisfaction outcomes in practice. Finally, our study focuses on developers who regularly use AI tools, which may skew results toward more favorable views of AI adoption.

**Customer Interviews and Observational Studies**

To augment our survey findings with qualitative learning, we've included insights collected from interviews and observational studies with professional developers and engineering leaders. In all, we interviewed 10 professional developers and 20 engineering leads about the impact LLMs were having on their software development teams and observed 23 experienced Java developers as they navigated both common (i.e., "accelerator") tasks as well as uncommon (i.e., "exploratory") tasks with GitHub Copilot. Finally, we also conducted interviews with a handful of engineers at Microsoft about their experiences using GitHub Copilot, specifically.

# Findings

## Widespread Adoption of AI Tools

The adoption of AI tools in software development is no longer a niche experiment, it's becoming the norm. While developers express concerns about AI being gimmicky, or potentially degrading their skills, that does not appear to be a major blocker to adoption.

In our survey, **75% of developers reported that they regularly use AI to complete their tasks**, while 25% indicated they do not.

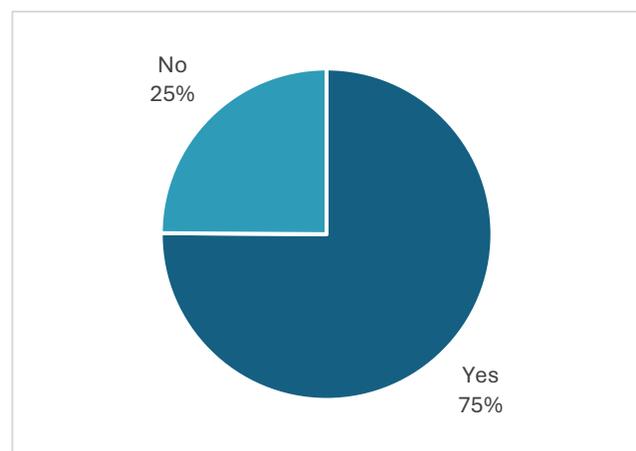

Figure 1: Response to question "*Do you regularly use AI tools in your role to help you accomplish your tasks*"

This widespread adoption signals a clear shift in how developers are integrating AI into their workflows. Among those who have adopted AI, usage is frequent. 64% of developers who use AI report using it at least once per week.

While seniority plays a role, its impact on adoption is relatively minor. The most experienced developers (7+ years of experience) are only 4% less likely to be daily users compared to those in their first three years. However, organizational support appears to be a far stronger driver of AI adoption. Developers who say their organization actively advocates for AI adoption are seven times more likely to be daily users than those who feel their company does not.

## Perceived Productivity Gains

Yet, adoption alone doesn't tell the full story. Are these tools actually making developers more productive? Are some aspects of work seeing more benefits than others? And what role does team-wide adoption play in maximizing impact?

The following sections break down these questions, exploring not just how often AI is used, but how it's shaping individual and team productivity across the SPACE dimensions.

**90%**
Percentage of devs who regularly use AI that said that AI makes them more productive.

**80%**
Percentage of devs who regularly use AI that said that they would be sad if they could no longer use it.

These results make it clear: Developers who use AI believe it has a meaningful impact on their "productivity". But what does productivity mean in this context? Productivity is more than just speed, it spans multiple dimensions, from efficiency and task throughput to job satisfaction and collaboration. To better understand AI's role, we examined its influence through the SPACE framework[3], which captures five key aspects of developer productivity:

- **Satisfaction (S)** – How happy and fulfilled developers feel in their work
- **Performance (P)** – Their ability to deliver meaningful business or customer value
- **Activity (A)** – The volume of code and tasks completed
- **Collaboration (C)** – How effectively they work with teammates
- **Efficiency (E)** – How smoothly and effectively they get work done

The following sections break down whether developers agree or disagree that AI is impacting their experiences across each of these dimensions.

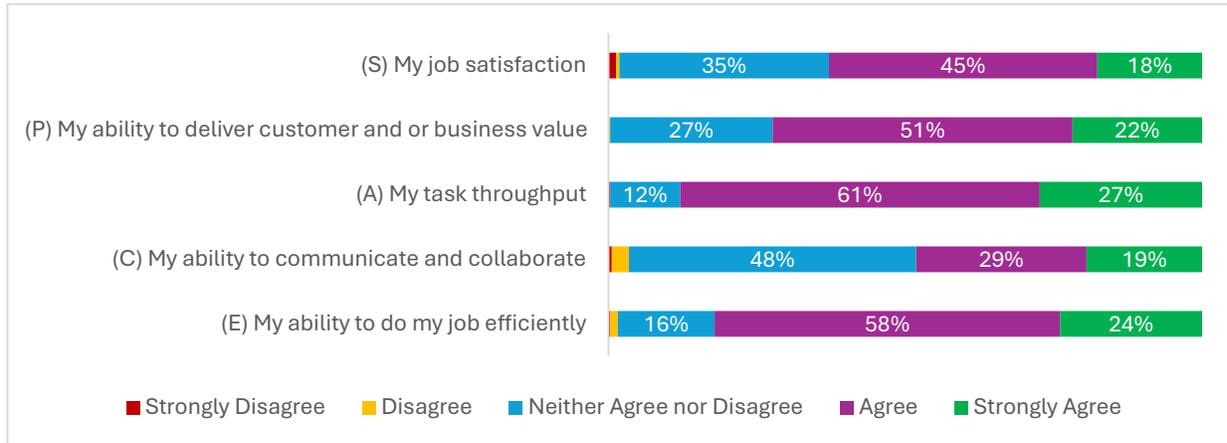

Figure 2: Percentage of developers who agree or disagree that AI has improved each aspect of their work.
Note: Due to the small number of developers who disagreed with these statements, disagreement segments are visible in the chart but not labeled.

The data leaves little room for doubt: For those developers using AI, they are seeing it improve their productivity across a wide range of dimensions. Less than 3% of respondents selected 'Disagree' or 'Strongly Disagree' for any item. A strong majority agree or strongly agree that it improves their task throughput (88%) and efficiency (82%), which aligns with previous findings about the efficacy of Copilot to decrease the time it takes to complete coding tasks[5].

Beyond just speed, 71% of AI adopters believe these tools help them deliver customer value, and 62% say AI enhances their job satisfaction. These results suggest that, for those who use AI, it's not just making them faster, it's helping them feel more effective and more satisfied with their work. Previous research has shown a positive relationship between productivity and satisfaction[6], and it would appear that these results lend agreement to that.

The impact on collaboration tells a more nuanced story. Unlike other productivity dimensions, fewer than half of AI adopters (48%) agreed that AI has improved their ability to collaborate with teammates. Notably, very few developers outright disagreed, but a significant portion remained neutral, suggesting that AI's role in team-based work is still evolving.

However, our qualitative findings reveal a more complex picture. Rather than decreasing collaboration, AI seems to be changing the nature of interactions within teams. The engineering managers we spoke to reported fewer interruptions within their teams, as developers relied less on colleagues for quick answers to coding questions. An executive leader at an online grocery delivery service even noted that developers were experiencing less "reputational damage" for having to interrupt teammates with simple coding questions.

At the same time, many leaders and developers did not feel that AI was diminishing meaningful discussions. Several interviewees claimed that the quality and depth of conversations, particularly between senior and junior developers, had actually improved. A VP of IT at a global supplier of technology and engineering systems explained that, because of ChatGPT, teams now spend more time "brainstorming about projects, ideas, and architectures" and less time discussing 'unnecessary' coding questions.

These findings suggest that while AI may not yet be seen as a strong direct enabler of collaboration, it is reshaping how collaboration happens. By reducing interruptions and shifting conversations toward higher-value discussions, AI may be indirectly improving team interactions in ways that traditional productivity metrics, and even developer self-assessments, don't fully capture. As teams continue to evolve their use of AI, its role in enhancing collaboration may become more visible and more measurable over time.

## Task Complexity and AI's Limits

While developers overwhelmingly believe that AI improves productivity across multiple dimensions, the type of work matters. Not all tasks are created equal, and this distinction is critical when evaluating AI's true impact. Developers who regularly use AI report that it excels at handling mundane, repetitive work, but struggles with more complex or novel challenges.

Here is how they described it, in their own words:

| "It takes care of so much **tedious** work!" | "I'm still stuck solving all the **hard problems**" |
| --- | --- |
| "I've been able to hand-off fundamentally **simple but repetitive** [tasks] to AI" | "I don't find [AI tools] particularly useful for **complex** situation" |
| "It's mostly a time-saver for the cases where I'm doing something **repetitive**" | "It definitely isn't perfect, especially for anything **complex**" |

Figure 3: Response to question "In your own words, how has the use of AI tools impacted your productivity?"

It's not just the nature of tasks that shapes AI's impact, how we use AI also matters. Developers who incorporate AI into their workflow more frequently may refine their approach, learning when and how to leverage it most effectively. Early interactions with AI might be limited to basic code suggestions, but over time, developers may integrate it into more nuanced problem-solving, documentation, and debugging efforts. As their familiarity grows, so too does AI's perceived value. The data reflects this: developers who use AI more often tend to report stronger productivity benefits.

## Frequency of Use Shapes Perceived Value

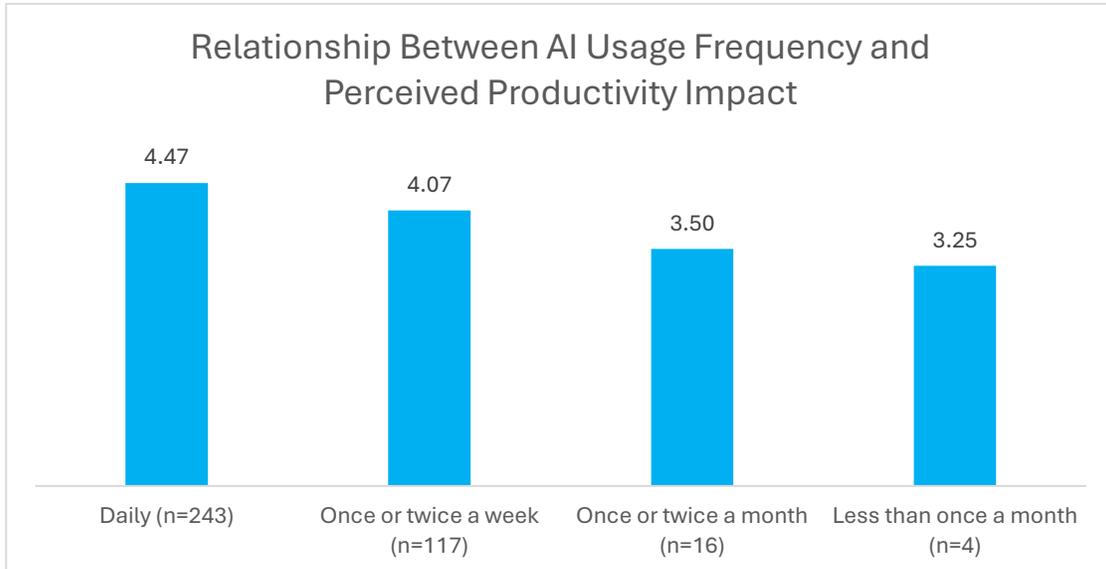

Figure 4: Response to question "Which best describes how often you use AI tools in your day-to-day work?" compared to the average agreement rating to the question "Using AI helps me to be more productive"

The data suggests a positive relationship between AI usage frequency and how strongly developers believe it improves their productivity. However, this trend alone does not imply causation. Due to relatively small sample sizes for less frequent AI users, the differences are not statistically significant, meaning they could have arisen by chance. The direction of the relationship also remains unclear: Does frequent AI use lead to greater productivity, or are developers who already see value in AI simply more likely to use it more often?

Our observational studies of professional Java developers provide additional insights. Success with AI depended on three key factors:

1) Task complexity
2) Developer skill
3) Familiarity with AI.

AI worked well for routine tasks but struggled with novel problems. More experienced developers evaluated AI-generated code more effectively, and those familiar with how LLMs work, were better at refining prompts and integrating AI into workflows.

Developers engage in a continuous loop when working with AI: formulating the problem, determining inputs, evaluating AI suggestions, and deciding on implementation.

Breakdowns at any step—misunderstanding the task, unclear prompts, or lack of trust in AI outputs—can lead to frustration and wasted effort. Many developers say working with AI requires new skills or forces them to engage skills they don't often use. A Distinguished Engineer at Microsoft described this learning curve:

> *"[GitHub Copilot Chat] took me a long time to use because I felt like Chat is for when I'm not in [coding] flow. That was not a muscle that I had at all."*

## Team-Wide Adoption Amplifies Impact

But AI's impact isn't just about how we use it; it's also shaped by the people around us. The way AI is adopted within a team can influence how individuals perceive its value, whether through shared best practices, cultural norms, or simply seeing colleagues benefit from its use. The next section explores how team-wide AI adoption affects perceptions of productivity.

| *"How many developers in your team use AI tools in their day-to-day role?"* | *"I feel like **my team** is productive at work"* | *"Using AI helps **me** to be more productive"* |
|---|---|---|
| All | 94% agree | 4.66 |
| Most | 86% agree | 4.35 |
| Some | 79% agree | 4.18 |

Figure 5 Responses to the question "*How many developers in your team use AI tools in their day-to-day role'*" compared against (1) the percentage of respondents who agree with the statement "*I feel like my team is productive at work*" and (2) the average agreement rating to "*Using AI helps me to be more productive' on a 5-point scale*."

As shown in figure 5, developers working on teams where AI is widely adopted are more likely to view their team as productive. This follows, logically, from the strong agreement developers have that AI makes themselves more productive. It stands to reason that the more of their peers who use this productivity-boosting tool, the more productive the team will be. This raises an important question about whether there will be growing social pressure to adopt AI, if the absence of adopting it will make individuals seem less productive in the eyes of their peers.

One possibility is that as AI becomes the norm within a team, those who don't use it may start to feel, or be perceived as, less productive, even if their work doesn't directly benefit from AI-powered assistance. If AI adoption is socially reinforced, developers may feel pressure to integrate it into their workflow, whether or not they see immediate value. Over time, this could shift team norms, making AI tools a default expectation rather than an optional enhancement.

However, the data also suggests that AI's benefits may extend beyond individual use through shared learning. Developers on teams with higher AI adoption don't just rate their team as more productive, they also report stronger personal agreement that AI makes themselves more productive. This could indicate that as AI becomes a standard tool within a team, developers gain exposure to best practices, workflow optimizations, and peer guidance that help them get more value from it.

This is particularly important given that AI's usefulness isn't universal across all tasks. Some developers may struggle to see the benefits initially, especially if their work requires deep problem-solving rather than automation of repetitive tasks. In a highly AI-adopted environment, these developers may learn from their peers about where AI is most effective, allowing them to refine their usage and ultimately see greater gains.

AI's impact on developer productivity is not just a function of the tool itself, but of how it is adopted, shared, and reinforced within teams. While individual usage plays a role, broader team adoption can shape perceptions, drive learning, and even create social expectations around AI's effectiveness. These findings highlight that measuring AI's value requires looking beyond just individual gain, its influence extends through the collaborative dynamics of teams and organizations. As AI continues to evolve, understanding these interactions will be key to maximizing its potential in software development.

## Practical Strategies for AI Impact

AI adoption in software development is accelerating, but how it is integrated and supported within teams and organizations plays a critical role in its effectiveness. While individual developers are finding value in AI tools, their impact is amplified when teams establish best practices, shared learning, and structured support systems. Below are key takeaways for teams, organizations, and researchers looking to maximize AI's potential.

**For Teams: Foster Best Practices and Encourage Team-Wide Adoption**

AI's effectiveness isn't just about whether an individual developer uses it, it thrives when adoption is supported across a team. As the data suggests, developers on AI-adopting teams report stronger productivity benefits, likely due to shared learning and collaborative improvements. To maximize AI's value:

- Create best practices documentation to help developers understand when and how to use AI effectively.

- Encourage knowledge sharing, such as team-wide discussions on AI-powered workflows or informal mentorship around AI usage. Appointing local champions can help drive adoption.

- Normalize AI adoption by embedding it into daily work processes rather than treating it as an optional tool.

Teams that openly discuss AI usage, share successes, and establish guidelines for responsible adoption will see stronger, more consistent benefits across developers.

**For Organizations: Invest in Training and Tooling to Drive AI Success**

Beyond individual teams, organizations play a crucial role in AI's effectiveness by shaping policy, training, and infrastructure. One of the most striking findings from this study is that developers who feel their organization actively supports AI adoption are seven times more likely to use it daily. This suggests that clear top-down encouragement, training, and access to AI tools can significantly drive usage. Organizations can:

- Provide structured AI training to help developers refine their usage patterns and integrate AI effectively into their workflows.
- Develop AI-friendly policies that encourage experimentation while addressing concerns about security, reliability, and ethical considerations.
- Ensure access to high-quality AI tools by evaluating internal and external options that best fit engineering needs.

Without these investments, driving AI adoption may remain challenging, limiting its potential benefits and creating unnecessary friction in development workflows.

**For Researchers: Investigate the Long-Term Impact of AI Adoption**

This study highlights strong correlations between AI usage and perceived productivity improvements, but many open questions remain. Future research should explore:

- Causality: Does frequent AI use drive greater productivity, or are more productive developers simply more likely to adopt AI?
- Task-Specific AI Impact: AI's effectiveness varies across tasks—where does it provide the most significant lift, and where does it fall short?
- The Social Dynamics of AI Adoption: How does peer influence shape AI adoption, and what role does social expectation play in driving usage?

Understanding these dynamics will help teams and organizations refine their AI strategies, ensuring that productivity gains are real, measurable, and sustainable over time.

# Conclusion

The integration of AI into software development is no longer an emerging trend, it is rapidly becoming a fundamental part of the developer experience. Our research shows that developers overwhelmingly believe AI enhances their productivity, efficiency, and satisfaction, yet its impact is far from uniform across all tasks and teams. While AI excels at automating routine and repetitive work, developers still find themselves tackling the most complex and strategic problems themselves.

At an individual level, familiarity and experience with AI influence how much benefit developers derive from these tools. Those who use AI more frequently tend to report greater productivity gains, though it remains unclear whether frequent use drives impact or if developers whose specific tasks benefit are best suited for AI, adopt it more often. Beyond individual usage, team-wide adoption plays a crucial role in shaping AI's effectiveness. Developers who work in AI-supportive environments not only rate their teams as more productive but also perceive stronger personal benefits, suggesting that peer learning and organizational support are key enablers of AI success.

Despite these promising trends, challenges remain. AI assistance requires new ways of thinking, and developers must build the skills necessary to formulate effective prompts, interpret AI suggestions, and evaluate their outputs critically. Organizations that invest in training, establish best practices, and embed AI into workflows will be better positioned to maximize its benefits.

Looking ahead, the conversation about AI in software development must evolve. The focus should shift from whether AI makes developers faster to how it reshapes the broader developer experience. By considering all dimensions of productivity—Satisfaction, Performance, Activity, Collaboration, and Efficiency (SPACE)—organizations can gain a more holistic view of AI's impact and ensure that its adoption leads to meaningful, sustainable improvements.

AI is not replacing developers; it is augmenting them. The challenge now is not just about measuring AI's effects, but understanding how to harness its potential in ways that truly empower developers, teams, and organizations.

## Acknowledgments

We would like to thank Lane Cooper, Monty Hammontree, Nancy Anderson and David Liu for their contributions to this research study. We would also like to thank all the study participants and research reviewers for their valuable feedback and insights.